\documentclass{article}
\usepackage[numbers, square, comma, sort&compress]{natbib} 

\usepackage{iclr2024_conference,times}

\usepackage{amsmath,amsfonts,bm}

\def\eqref#1{equation~\ref{#1}}

\def\1{\bm{1}}

\DeclareMathAlphabet{\mathsfit}{\encodingdefault}{\sfdefault}{m}{sl}
\SetMathAlphabet{\mathsfit}{bold}{\encodingdefault}{\sfdefault}{bx}{n}

\usepackage{hyperref}
\usepackage{url}
\usepackage{amsmath,amssymb,amsfonts}
\usepackage{algorithmic}
\usepackage{graphicx}
\usepackage{subcaption}
\usepackage{booktabs}
\usepackage{float}
\usepackage[format=plain, font=it]{caption}
\usepackage{textcomp}
\usepackage{xcolor}
\def\BibTeX{{\rm B\kern-.05em{\sc i\kern-.025em b}\kern-.08em
    T\kern-.1667em\lower.7ex\hbox{E}\kern-.125emX}}

\title{Luganda Speech Intent Recognition for IoT Applications}

\author{John Trevor Kasule \& Elvis Mugume\\
Department of Electrical and Computer Engineering\\
Makerere University\\
Kampala, Uganda \\
\texttt{\{john.kasule,elvis.mugume\}@students.mak.ac.ug} \\
\And
Sudi Murindanyi \\
Department of Computer Science, \\
Makerere University \\
Kampala, Uganda \\
\texttt{sudi.murindanyi@students.mak.ac.ug} \\
\AND
Andrew Katumba
\thanks{ Corresponding Author} \\
Department of Electrical and Computer Engineering\\
Makerere University\\
Kampala, Uganda \\
\texttt{andrew.katumba@mak.ac.ug}
}

\iclrfinalcopy 
\begin{document}

\maketitle

\begin{abstract}
The advent of Internet of Things (IoT) technology has generated massive interest in voice-controlled smart homes. While many voice-controlled smart home systems are designed to understand and support widely spoken languages like English, speakers of low-resource languages like Luganda may need more support. This research project aimed to develop a Luganda speech intent classification system for IoT applications to integrate local languages into smart home environments. The project uses hardware components such as Raspberry Pi, Wio Terminal, and ESP32 nodes as microcontrollers. The Raspberry Pi processes Luganda voice commands, the Wio Terminal is a display device, and the ESP32 nodes control the IoT devices. The ultimate objective of this work was to enable voice control using Luganda, which was accomplished through a natural language processing (NLP) model deployed on the Raspberry Pi. The NLP model utilized Mel Frequency Cepstral Coefficients (MFCCs) as acoustic features and a Convolutional Neural Network (Conv2D) architecture for speech intent classification. A dataset of Luganda voice commands was curated for this purpose and this has been made open-source. This work addresses the localization challenges and linguistic diversity in IoT applications by incorporating Luganda voice commands, enabling users to interact with smart home devices without English proficiency, especially in regions where local languages are predominant.
\end{abstract}

\section{Introduction}
Rapid technological advancements have led to the proliferation of smart devices and the Internet of Things (IoT) in various domains. Smart homes, in particular, have gained significant popularity, offering convenience, efficiency, and enhanced control over home appliances and systems \citep{dinushika2019speech}. Voice-controlled smart home systems have emerged as a convenient and intuitive way to interact with these devices using voice commands \citep{michaelconversationalinterface}. These systems respond to voice commands and must extract the user's intent from the command to be effective, known as intent recognition. The intent is something the user wants to do, which may be something like turning on the lights, decreasing the volume, or increasing the fan's speed. Intent recognition, a sub-task of Spoken Language Understanding, aims to extract the user's aim in a voice command\citep{desot2019towards}. However, most existing voice-controlled smart home systems are primarily designed to support notable languages such as English, leaving users of low-resource languages disadvantaged. In countries like Uganda, where English is not widely spoken, the existing systems fail to cater to the needs and preferences of the local population \citep{Building_Text_and_Speech}, \citep{murindanyi2023explainable}. This work addresses this language barrier by developing a speech intent classification system for Luganda, a prominent Bantu language and the mother tongue of Uganda's largest ethnic group, the Baganda \citep{explainable_MT}. Additionally, it leverages the power of artificial intelligence and machine learning techniques to enable Ugandan users to interact with smart home devices using a local language, enhancing user experience and accessibility. We recognize the importance of data collection, preprocessing, and model development to achieve accurate and reliable speech intent classification. By collecting a substantial amount of data from diverse participants and employing techniques such as the use of convolutional neural networks (CNNs), the work aims to develop a robust and efficient model capable of accurately classifying spoken commands related to various smart home devices and functions. The work explores post-training model quantization techniques to choose an appropriate model size for deployment on edge devices. Furthermore, the work extends its scope to include integrating IoT devices, allowing the classified speech intents to be translated into actionable commands that control the corresponding smart devices.  This integration involves using the MQTT\footnote{https://mqtt.org/} (Message Queuing Telemetry Transport) protocol, which enables seamless communication between the Raspberry Pi and other IoT devices, such as the Wio terminal\footnote{https://www.seeedstudio.com/Wio-Terminal-p-4509.html} and ESP32 boards. By empowering users to interact with their smart devices in their native language, the work aims to improve accessibility, usability, and overall user satisfaction in smart home automation.

The remaining sections are organized as follows: Section Two discusses related work presented in private resources, Section Three discusses paper contributions, and Section Four outlines the methodologies and approaches used. Section five presents the results and the discussions, and section six concludes the paper and suggests future work.

\section{Related Work}
This section explores speech intent recognition in IoT applications, focusing on two key methodologies: direct speech-to-intent and speech-to-text-to-intent. The direct method, highlighted by the work of \citeauthor{buddhika2018domain} \citep{buddhika2018domain}, utilizes Mel-Frequency Cepstral Coefficients (MFCCs) and Mel spectrograms to accurately classify speech intent in Sinhala using a neural network, demonstrating the importance of audio feature engineering. Complementary research by \citeauthor{commands_CNN} \citep{commands_CNN} shows that CNNs excel in speech command classification, outperforming traditional neural network models by effectively processing varied speech commands. Alternatively, the speech-to-text-to-intent approach, as improved by multiple pre-trained Automatic Speech Recognition (ASR) models presented in different papers, for example, the work done by \citeauthor{9906230} \citep{9906230}, transcribes speech-to-text before intent extraction, addressing the challenge of error propagation from ASR to  Natural Language Understanding (NLU). This method enhances recognition in limited-resource settings, showcasing advancements in speech intent recognition technology that cater to diverse IoT applications.  The end-to-end method, proposed by \citeauthor{luren} \citep{luren}, aims for direct speech-to-intent mapping, requiring significant domain-specific data but offering a unified solution that circumvents ASR inaccuracies. The choice between these approaches depends on the specific application and available resources.

Several valuable resources have emerged in the landscape of spoken language understanding (SLU) datasets, each contributing to the development and evaluation of SLU systems. Among these is the Air Travel Information System (ATIS) \cite{hemphill-etal-1990-atis}, an audio dataset with 17 intents related to air travel planning.  The Snips SLU Dataset is another publicly available asset encompassing English and French languages with only about 7 intents. Recent efforts have been made to curate more extensive datasets for end-to-end SLU tasks, such as the Fluent Speech Commands (FSC) \cite{luren} with 31 intents, A Spoken Language Understanding Resource Package (SLURP) \cite{bastianelli2020slurp} with about 18 domains, and Spoken Task Oriented Semantic Parsing (STOP) \cite{tomasello2022stop} datasets. These datasets have significantly contributed to the advancement of SLU technology, particularly in the personal assistant domain. 

Work has also been done in the domain of Speech Intent recognition for Smart homes. In \cite{3smart_home_prototype}, \citeauthor{3smart_home_prototype} introduce a smart home automation prototype that utilizes natural language processing and machine learning, enabling voice-controlled device operation, including lights, doors, and fans, through an interactive humanoid central processing unit. \citeauthor{13kumar_nlp_iot_smart_homes}  \cite{13kumar_nlp_iot_smart_homes}  present a Smart Home Automation system employing Natural Language Processing (NLP) and IoT cloud solutions for remote and secure smart home control. This infrastructure uses communication protocols to integrate user sound or voice commands transmitted to various appliances. The paper provides a prototype illustrating voice-controlled devices and sensors, offering an efficient and secure approach to smart home control through NLP and IoT cloud solutions. In \cite{23Reshamwala2021}, \citeauthor{23Reshamwala2021} proposes an IoT-based Home Automation system utilizing 2 hardware devices, sensors, and software applications like Google Firebase for comprehensive smart home control. This system allows remote appliance control via a mobile application from anywhere globally and incorporates voice commands through Google's Voice Assistant system.

\section{Paper Contribution}
Our work addresses gaps in Luganda intent classification for IoT device control, including the absence of a Luganda voice command dataset and a preliminary classification model.  We develop an IoT application that accurately recognizes Luganda speech commands focusing on regions where Luganda is widely spoken. Unlike existing papers that often focus on cloud-based IoT solutions, our approach is localized, deploying NLP models on Raspberry Pi and Wio Terminal nodes for efficient smart home automation, even in areas with limited internet connectivity. Furthermore, our work offers a comprehensive IoT implementation, covering model design, deployment, and integration with IoT devices, addressing gaps in prior literature. We have curated a Luganda voice command dataset with 20 distinct intents, and this dataset has been made open source and can be accessed at this \href{https://doi.org/10.7910/DVN/RPU3SM}{DOI}. Our integration of NLP and IoT technologies on Raspberry Pi and ESP32 nodes provides an intuitive means of controlling devices using local language commands. While delivering cutting-edge performance in spoken language understanding tasks, the speech-to-text-to-intent methodology employs Automated Speech Recognition (ASR) models that demand substantial computational resources despite being readily available. This requirement can pose significant challenges when deploying such models on resource-constrained devices like the Raspberry Pi and Wio Terminal. This limitation highlights the need for a tailored approach that considers both the computational constraints of the hardware and the specific requirements of the Luganda language domain. Moreover, the performance of these ASR models might not be optimal for applications within the Luganda speech intent domain, particularly in the context of smart home automation. Therefore inspired by previous work, we focused on the speech-to-intent approach, employing MFCCs as acoustic features and CNNs to enhance the model's performance in real-time IoT environments. Augmentation techniques improve model robustness, and model quantization techniques optimize memory usage for successful edge device deployment, aligning with our work's objectives.

\section{Methodology}

\subsection{Data Collection}

The process of collecting data for this research project involved two main steps. First, a list of utterances for each intent was gathered. Then, the dataset was collected through crowd-sourcing, which involved recruiting multiple individuals to record the utterances. The intents were identified on 2 slots: \textit{{action, device}}.  The smart home devices and their associated intents considered during the dataset collection are in the table \ref{tab:intents-actions}.
\begin{table}[h!]
\centering
\caption{Intents and Actions for Home Automation Devices}
\label{tab:intents-actions}
\begin{tabular}{|l|l|}
\hline
\textbf{Device} & \textbf{Actions} \\
\hline
Lights, Alarm, TV, Fridge, Camera & - On, - Off\\
\hline
Door & - open, - close\\
\hline
Fan & - On, - Off, - Increase speed, - Decrease speed \\
\hline
Speaker & - On, - Off, - Increase volume, - Decrease volume \\
\hline
\end{tabular}
\end{table}
\\Each of the 20 intents may have multiple transcriptions to account for variations in natural language expressions. For instance, the "lights on" intent in Luganda might include transcriptions like "saako ettaala," "ntelaako ettaala," "ettaala gyitekeeko," and more. The data was collected using smartphones, capturing audio recordings in various formats such as .wav, .aac, and .m4a. All audio files were converted to the 16KHz .wav format for consistency using ffmpeg\footnote{https://ffmpeg.org/}, an audio processing tool.\\ A total of 10,200 audio files were collected from 81 participants (43 males and 38 females, aged 20 to 25), divided into three sets: 8,127 for training, 1,068 for testing, and 1,005 for validation. Special attention was paid to the test set, ensuring that audio files used for this set were recorded from 7 (4 male and 3 female) speakers, and none appeared in the training set. This was done to simulate real-world scenarios where the model may face unseen speakers during deployment, thus assessing the model's robustness. This dataset has been made publicly available on Havard Dataverse\footnote{https://dataverse.harvard.edu/} at this \href{https://doi.org/10.7910/DVN/RPU3SM}{DOI} as the \textit{Luganda Voice Commands Dataset}.


\subsection{Data preprocessing}
Data preprocessing involved audio signal processing, data augmentation, and feature extraction, with separate experiments conducted for the Wio Terminal and Raspberry Pi. The data augmentation process was implemented to expand the training dataset, providing new variations and realistic scenarios for the model to learn from. This function took the original signal as input and returned the original signal along with two augmented versions: one with added white noise and another with pitch scaling applied. MFCC features were then extracted from raw audio. 
Experiments involved feature extraction from both augmented and unaugmented audio data. For the Wio Terminal, 10 features were extracted to assess the impact of feature dimensionality, while for the Raspberry Pi, two feature sets were used: 10 MFCCs and 13 MFCCs.

\subsection{Model development}\label{AA}
\begin{figure*}[htbp]
    \centering
        \includegraphics[scale=0.40, keepaspectratio]{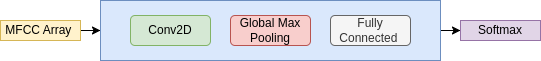}

    \caption{Model Architecture}
    \label{fig:model_architecture}
\end{figure*}

This study employed CNNs to develop the Luganda speech intent classification model using the extracted MFCC features. The model architecture, hyperparameters, and training process were fine-tuned to maximize the validation accuracy using an EarlyCallback callback with a patience of 20. The model architecture was as shown in figure \ref{fig:model_architecture}.\\
Two variations of CNN model architectures were used. The first included multiple Conv2D layers followed by BatchNormalization and Max-Pooling layers. It included a Global Max Pooling, dropout, and dense layers. A softmax activation function was applied at the classification layer, transforming the previous dense layer's output into a probability distribution over the different classes. In a modified architecture, the Flattening layer replaced the Global Max Pooling layer, reshaping the output into a 1-dimensional vector before being fed into the subsequent layers. Both architectures were trained and evaluated using the respective sets of audio features.
The choice of models for the Wio Terminal and Raspberry Pi was based on experiments detailed in the next subsection (\ref{experiments}).

\subsection{Experiments} \label{experiments}
We conducted experiments to identify the best model architecture and data augmentation techniques for deployment on the Wio Terminal and Raspberry Pi. Four experiments (A-D) were carried out for the Wio Terminal and eight (1-8) for the Raspberry Pi. Experiments A and B for the Wio Terminal and 1 and 2 for the Raspberry Pi assessed models trained with and without data augmentation, using 10 MFCCs with a Global Max Pooling layer. Experiments C and D for the Wio Terminal and 3 and 4 for the Raspberry Pi replicated these but used a Flattening layer instead of Global Max Pooling. Experiments 5 and 6 for the Raspberry Pi explored using 13 MFCCs, with and without data augmentation. Finally, Experiments 7 and 8 were examined using a Flattening layer with 13 MFCCs, again with and without data augmentation. The optimal model was chosen for quantization for each deployment mode.

\subsection{Model Deployment}
The best models from the quantization process for deployment on the Wio Terminal and Raspberry Pi were deployed on the corresponding devices.

\subsubsection{Deployment on the Wio Terminal}
The model was converted to TensorFlow Lite format with int8 quantization and transformed into a .h file for Arduino sketches. The Wio Terminal's microphone captured the audio, and MFCC calculation was applied to extract relevant features. Inference was done on the features and results were displayed on the screen. However, due to the board's processing limitations, it could not support both the model and WiFi capabilities, so we turned to Raspberry Pi to proceed with IoT integration goals.


\subsubsection{Deployment on Raspberry Pi} \label{dep_rasp}
We deployed the intent classification model on a Raspberry Pi for real-time inference and integration into the smart home system. Audio snippets were captured at 16000KHz using reSpeaker 2-Mics Pi HAT, followed by extraction of features for speech intent classification. Upon classification, commands were broadcast to IoT devices, enabling voice-controlled device operation in Luganda.


\subsection{IoT Integration}
The Luganda speech intent classification system was integrated with the IoT devices in the smart home environment. The process involved connecting multiple devices through a Wi-Fi connection provided by a smartphone. The Raspberry Pi was the central hub for processing and distributing the inferred commands to the respective IoT devices. The MQTT protocol was utilized for efficient and reliable message transmission. Upon inferring the speech intent, the Raspberry Pi sent the corresponding command to the IoT devices using the MQTT protocol. It published the command to the topic called \textit{rpi/broadcast}, to which the ESP32 boards and the Wio Terminal subscribed. The Raspberry Pi also inferred the confidence level with which the prediction was made. If the confidence level is less than 75\%, it uses a unique client ID to identify itself on the network and uses callbacks to receive MQTT messages. Upon receiving a command, the respective board interprets the message and executes commands on its screen, providing real-time feedback to the user.
\begin{figure}[htbp]
    \centering
    \includegraphics[scale=0.10, angle=90]{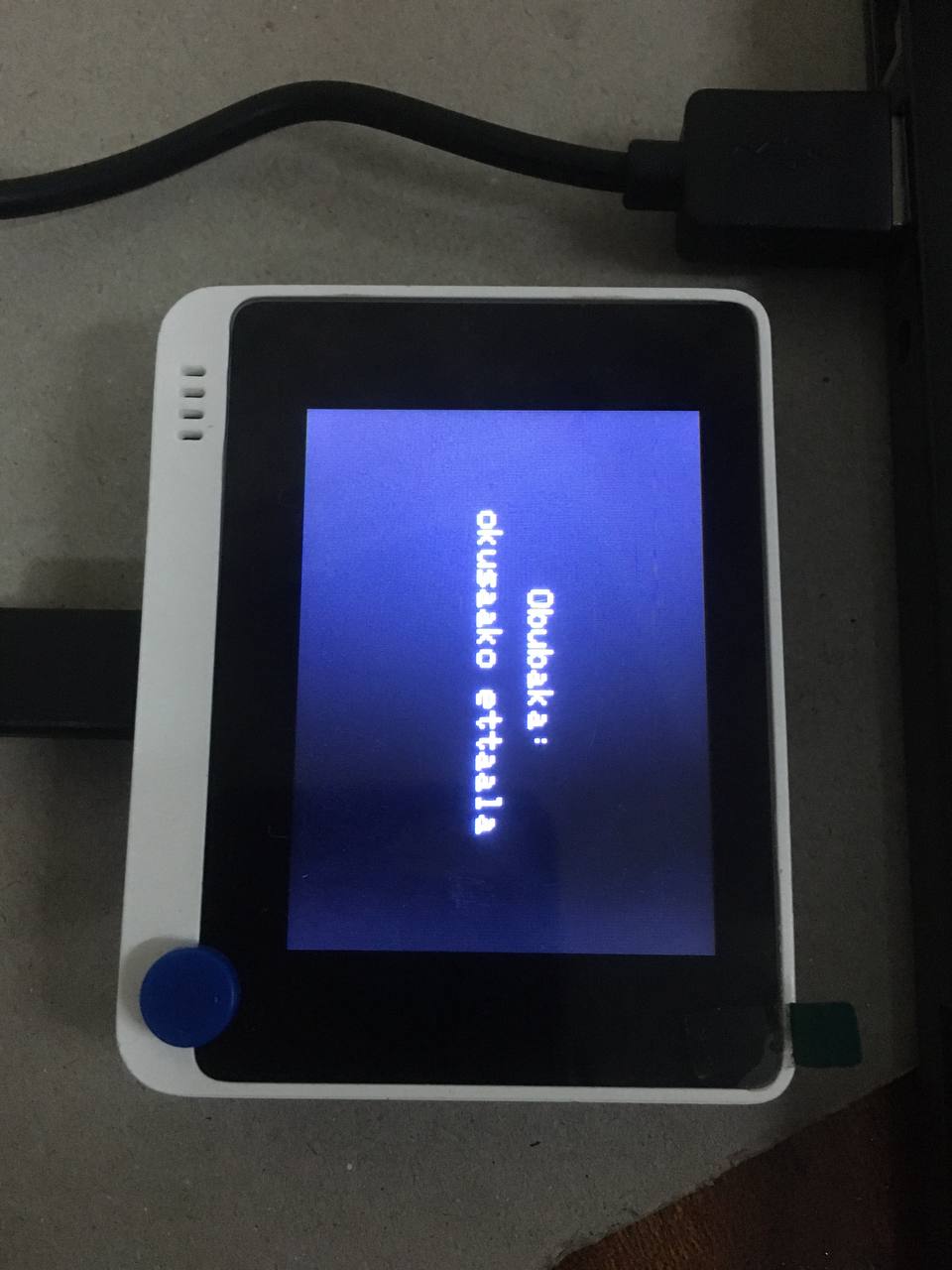}
    \caption{Wio Terminal display of inference}
    \label{fig:wio_mqtt}
\end{figure}
Each ESP32 board uses a unique client ID to identify itself on the network and uses callbacks to receive MQTT messages. Upon receiving a command, the respective board interprets the message and executes the corresponding action on the IoT device. 
During the demonstration of the smart home, we used some simple devices to showcase all of the items in the house. These devices included an alarm: \textit{buzzer}, a door:\textit{servo motor},  fan: \textit{dc motor}, lights: \textit{solar bulbs}, camera:\textit{blue LED}, television:\textit{yellow LED} and  fridge:\textit{red LED}.\\
An evaluation was done to assess the performance of the Luganda speech intent classification system. The system performed well in low-noise environments, with the model's accurate predictions and higher confidence levels. It demonstrated fairness in accuracy across different speakers' genders. However, it faced challenges with non-fluent speakers due to variations in pronunciation and accents.


\section{Results}
\subsection{Wio Terminal}
The methodology involved training the model to deploy it on the Wio Terminal, which involved 4 experiments as presented in section \ref{experiments}. The results of this training phase are presented in Table \ref{tab:results_wio}, which showcases the outcomes of Experiments A to D. Table \ref{tab:results_quant_wio} shows results obtained after quantizing the model for deployment on the Wio Terminal.
\begin{table*}[htbp]
    \centering
    \caption{Results of Data Extraction and Model Development Experiments for the Wio Terminal}
    \label{tab:results_wio}
    \begin{tabular}{l|clcccc}
        \toprule
        \textbf{Exp} & \textbf{MFCCs} & \textbf{Augmentation} & \textbf{Architecture} & \textbf{Accuracy(\%)} & \textbf{Test Loss} & \textbf{Parameters}\\
        \midrule
        A & 10 & Without & Global Max Pooling & 84.64 & 0.6105 & 45,524\\
        B & 10 & With & Global Max Pooling & 87.08 & 0.5252 & 45,524\\
        C & 10 & Without & Flattening &  72.10 & 1.2226  & 324,052\\
        D & 10 & With & Flattening & 82.02 & 0.8250 & 324,052\\
        \bottomrule
    \end{tabular}
\end{table*}


\begin{table*}[htbp]
    \centering
    \caption{Results of Model Quantisation: Wio Terminal}
    \label{tab:results_quant_wio}
    \begin{tabular}{lcccc}
        \toprule
        \textbf{Experiment} & \textbf{size(KB)} & \textbf{Model Accuracy} & \textbf{Accuarcy Tradeoff} & \textbf{Reduction\%} \\
        \midrule
        .h5 model & 654.75 & 87.08 & 0 & 0\\
        Default & 183.762 & 87.08 & 0 & 71.94\\
        Float 16 & 99.051 & 87.08 & 0 & 84.88 \\
        Int 8 & 59.242 & 84.55 & - 2.53 & 90.954 \\
        Size & 58.703 & 87.36 & +0.28 & 91.035\\

        \bottomrule
    \end{tabular}
\end{table*}

Experiment B, which utilized augmentation with 10 MFCCs and the Global Max Pooling architecture, achieved the highest accuracy of 87.08\% and a low loss value of 0.5252. These results demonstrated that this configuration effectively captured and generalized data patterns, making it suitable for real-world scenarios. Consequently, we selected the model from Experiment B for quantization, expecting it to maintain high accuracy while reducing model size and computational requirements. The quantization experiments on the Wio Terminal revealed a promising trade-off between model size and accuracy. The int8 quantization method was particularly effective, reducing the model size by 90.954\% while maintaining reasonable accuracy at 84.55\%. This method allowed efficient resource management, making it a suitable choice for deployment on the Wio Terminal.


\subsection{Raspberry Pi}
The Raspberry Pi model training results, comprising Experiments 1 to 8, are summarized in Table \ref{tab:results}. Additionally, Table \ref{tab:results_quant} illustrates the effects of various quantization techniques on both model size and accuracy.\\
The results show that models with Global Max Pooling consistently outperformed those with Flattening layers, with Experiment 2 achieving 87.08\% accuracy compared to Experiment 4's 80.43\% with the same number of features. Data augmentation also consistently improved accuracy. For instance, Experiment 2 with augmentation achieved 87.08\% accuracy, while Experiment 1 without reached 84.64\% on the same set of features. It is also observed that models with 13 MFCCs generally outperformed those with 10. Experiment 6, with 13 MFCCs and augmentation, scored 87.36\%, whereas Experiment 2, with 10 MFCCs and augmentation, reached 87.08\%.
It can, therefore, be concluded that models with Global Max Pooling, data augmentation, and 13 MFCCs give the best configuration for a good performance. \\
These insights guided our approach to \textit{Slot filling}, where the Global Max Pooling layer and 13 MFCCs with data augmentation were used to classify the model performance based on the \textit{action} and \textit{object} slots. The model achieved 94.29\% accuracy for the Action slot and 91.85\% for the Object slot, demonstrating high accuracy in classifying actions and objects. 
\begin{table*}[htbp]
    \centering
    \caption{Results of Data Extraction and Model Development Experiments: Raspberry Pi}
    \begin{tabular}{l|cccccc}
        \toprule
        \textbf{Exp.} & \textbf{MFCCs} & \textbf{Augmentation} & \textbf{Architecture} & \textbf{Accuracy(\%)} & \textbf{Test Loss} & \textbf{Parameters}\\
        \midrule
        1 & 10 & Without & Global Max Pooling & 84.64 & 0.7150 & 45,524\\
        2 & 10 & With & Global Max Pooling & 87.08 & 0.4709 & 45,524\\
        3 & 10 & Without & Flattening &  80.15 & 0.7582  & 274,900\\
        4 & 10 & With & Flattening & 80.43 & 0.9648 & 274,900\\
        \midrule
        5 & 13 & Without & Global Max Pooling & 83.80 & 0.7144 & 45,524\\
        6 & 13 & With & Global Max Pooling & 87.36 & 0.6486 & 45,524 \\
        7 & 13 & Without & Flattening &  83.05 & 0.8888 & 274,900\\
        8 & 13 & With & Flattening & 80.99 & 1.1151 & 274,900 \\
        \bottomrule
    \end{tabular}
    
    \label{tab:results}
\end{table*}
\begin{figure}[htbp]
    \centering
\includegraphics[scale=0.75,width = 0.75\linewidth, height=0.75\textheight, keepaspectratio]{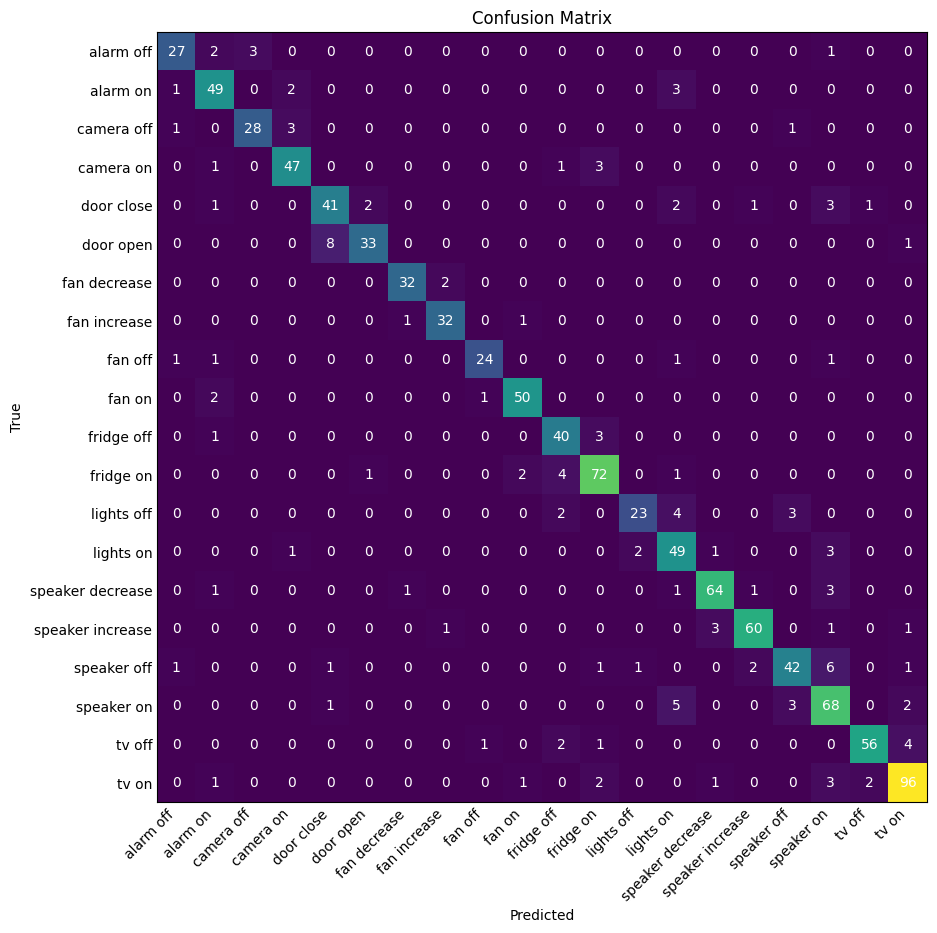}
    \caption{Confusion Matrix for Experiment 6}
    \label{fig:cm_6}
\end{figure}
\begin{table*}[htb!]
    \centering
    \caption{Results of Model Optimization}
    \label{tab:results_quant}
    \begin{tabular}{lcccc}
        \toprule
        \textbf{Experiment} & \textbf{Model size(KB)} & \textbf{Accuracy} & \textbf{Accuracy change} & \textbf{Reduction\%} \\
        \midrule
        .h5 model & 647.5 & 87.36 & 0 & 0\\
        Default & 183.859 & 87.36 & 0 & 71.60\\
        Float 16 & 99.211 & 87.36 & 0 & 84.67 \\
        Size & 58.805 & 87.73 & +0.37 & 90.91\\

        \bottomrule
    \end{tabular}
\end{table*}\\
Consequently, the model from Experiment 6 and the slot-filling model were selected for further analysis. In real-time, slot filling can create new intent combinations, such as “lights decrease” from the object and slot outputs, "lights" and "decrease," respectively, which may not exist in the original dataset or be catered for in hardware integration. Experiment 6, on the other hand, used a single output approach, achieving an accuracy of 87.36\% and a low loss value of 0.6486. It allowed for effective monitoring of confidence levels, ensuring consistency and reliability in real-time applications, and prevented the creation of new intents at deployment. Given its strong performance and advantages, Experiment 6 was chosen for further analysis in the context of model quantization, and results are shown in Table \ref{tab:results_quant}. The default quantization technique reduced the model size to 183.859 KB, maintaining an accuracy of 87.36\%. The Float16 technique further reduced the size to 99.211 KB without accuracy loss. The size optimization technique resulted in the smallest model size of 58.805 KB, a 90.91\% reduction, and slightly improved accuracy to 87.73\%. Given the Raspberry Pi's resource and memory constraints, the size-optimized model was chosen for deployment due to its balance of size reduction and accuracy improvement. This model was deployed for real-time inference on the Raspberry Pi, ensuring efficient and reliable predictions with reasonable accuracy.

\section*{Conclusion}
This research significantly advances Luganda speech intent recognition for IoT applications. It introduces a crucial dataset of Luganda voice commands and employs MFCCs for effective feature extraction. Various CNN architectures and data augmentation techniques enhance the model's accuracy. Furthermore, the model was optimized for deployment on resource-constrained devices through quantization, ensuring high accuracy and reliability for seamless integration with IoT devices. However, our ongoing efforts are focused on developing a real-time hot word detection system. This addition will enhance user interactions by enabling hands-free voice commands, making IoT devices even more accessible and user-friendly. This work contributes to the IoT experience for Luganda-speaking users, enhancing user experiences and accessibility in smart homes. It bridges language barriers, enabling integrated IoT in diverse contexts. We have made our Luganda voice command dataset open source, which could significantly aid in developing similar technologies and ensure the reproducibility of results. Future work can explore multi-language support and advanced machine-learning techniques for further improvements. We also plan to integrate our work with established frameworks for low-resource Spoken Language Understanding, such as the ESPNet framework, to ensure that our research on Luganda does not remain an isolated endeavour but becomes part of a broader, community-driven initiative.

\subsubsection*{Acknowledgments}
We extend our deepest gratitude to the Marconi Research and Innovations Laboratory\footnote{https://marconilab.org/} at Makerere University for their generous funding and the provision of resources for our project.

\bibliography{iclr2024_conference}

\begin{thebibliography}{16}
\providecommand{\natexlab}[1]{#1}
\providecommand{\url}[1]{\texttt{#1}}
\expandafter\ifx\csname urlstyle\endcsname\relax
  \providecommand{\doi}[1]{doi: #1}\else
  \providecommand{\doi}{doi: \begingroup \urlstyle{rm}\Url}\fi

\bibitem[Babirye et~al.(2022)Babirye, Nakatumba-Nabende, Katumba, Ogwang, Francis, Mukiibi, Ssentanda, Wanzare, and David]{Building_Text_and_Speech}
Claire Babirye, Joyce Nakatumba-Nabende, Andrew Katumba, Ronald Ogwang, Jeremy~Tusubira Francis, Jonathan Mukiibi, Medadi Ssentanda, Lilian~D Wanzare, and Davis David.
\newblock Building text and speech datasets for low resourced languages: A case of languages in east africa.
\newblock In \emph{3rd Workshop on African Natural Language Processing}, 2022.
\newblock URL \url{https://openreview.net/forum?id=SO-U99z4U-q}.

\bibitem[Bastianelli et~al.(2020)Bastianelli, Vanzo, Swietojanski, and Rieser]{bastianelli2020slurp}
Emanuele Bastianelli, Andrea Vanzo, Pawel Swietojanski, and Verena Rieser.
\newblock Slurp: A spoken language understanding resource package, 2020.

\bibitem[Buddhika et~al.(2018)Buddhika, Liyadipita, Nadeeshan, Witharana, Javasena, and Thayasivam]{buddhika2018domain}
D.~Buddhika, R.~Liyadipita, S.~Nadeeshan, H.~Witharana, S.~Javasena, and U.~Thayasivam.
\newblock Domain specific intent classification of sinhala speech data.
\newblock In \emph{2018 International Conference on Asian Language Processing (IALP)}, pp.\  197--202, Nov. 2018.
\newblock \doi{10.1109/IALP.2018.8629103}.

\bibitem[Desot et~al.(2019)Desot, Portet, and Vacher]{desot2019towards}
T.~Desot, F.~Portet, and M.~Vacher.
\newblock Towards end-to-end spoken intent recognition in smart home.
\newblock In \emph{2019 International Conference on Speech Technology and Human-Computer Dialogue (SpeD)}, pp.\  1--8, 2019.
\newblock \doi{10.1109/SPED.2019.8906584}.

\bibitem[Dinushika et~al.(2019)Dinushika, Kavmini, Abeyawardhana, Thayasivam, and Jayasena]{dinushika2019speech}
T.~Dinushika, L.~Kavmini, P.~Abeyawardhana, U.~Thayasivam, and S.~Jayasena.
\newblock Speech command classification system for sinhala language based on automatic speech recognition.
\newblock In \emph{2019 International Conference on Asian Language Processing (IALP)}, pp.\  205--210, 2019.
\newblock \doi{10.1109/IALP48816.2019.9037648}.

\bibitem[Hemphill et~al.(1990)Hemphill, Godfrey, and Doddington]{hemphill-etal-1990-atis}
Charles~T. Hemphill, John~J. Godfrey, and George~R. Doddington.
\newblock The {ATIS} spoken language systems pilot corpus.
\newblock In \emph{Speech and Natural Language: Proceedings of a Workshop Held at Hidden Valley, {P}ennsylvania, June 24-27,1990}, 1990.
\newblock URL \url{https://aclanthology.org/H90-1021}.

\bibitem[Kobusingye et~al.(2023)Kobusingye, Dorothy, Nakatumba-Nabende, and Marvin]{explainable_MT}
Belinda~Marion Kobusingye, Ankunda Dorothy, Joyce Nakatumba-Nabende, and Ggaliwango Marvin.
\newblock Explainable machine translation for intelligent e-learning of social studies.
\newblock In \emph{2023 7th International Conference on Trends in Electronics and Informatics (ICOEI)}, pp.\  1066--1072, 2023.
\newblock \doi{10.1109/ICOEI56765.2023.10125599}.

\bibitem[Kumar et~al.(2019)Kumar, Benedict, and Ajith]{13kumar_nlp_iot_smart_homes}
S.~Kumar, S.~Benedict, and S.~Ajith.
\newblock Application of natural language processing and iotcloud in smart homes.
\newblock In \emph{2019 2nd International Conference on Intelligent Communication and Computational Techniques (ICCT)}, pp.\  20--25, 2019.
\newblock \doi{10.1109/ICCT46177.2019.8969066}.

\bibitem[Lugosch et~al.(2019)Lugosch, Ravanelli, Ignoto, Tomar, and Bengio]{luren}
Loren Lugosch, Mirco Ravanelli, Patrick Ignoto, Vikrant~Singh Tomar, and Yoshua Bengio.
\newblock Speech model pre-training for end-to-end spoken language understanding, 2019.
\newblock URL \url{https://arxiv.org/abs/1904.03670}.

\bibitem[McTear et~al.(2016)McTear, Callejas, and Griol]{michaelconversationalinterface}
Michael McTear, Zoraida Callejas, and David Griol.
\newblock \emph{The conversational interface: Talking to smart devices}.
\newblock Springer International Publishing, January 2016.
\newblock ISBN 9783319329659.
\newblock \doi{10.1007/978-3-319-32967-3}.

\bibitem[M.K.M et~al.(2018)M.K.M, Kumar, Sharma, Pasha, and K.H.D]{3smart_home_prototype}
M.K.M, K.B.~Mukesh Kumar, L.~Sharma, M.Z.~Sayeed Pasha, and K.H.D.
\newblock An interactive voice controlled humanoid smart home prototype using concepts of natural language processing and machine learning.
\newblock In \emph{2018 3rd IEEE International Conference on Recent Trends in Electronics, Information \& Communication Technology (RTEICT)}, pp.\  1537--1546, 2018.
\newblock \doi{10.1109/RTEICT42901.2018.9012359}.

\bibitem[Mohamed \& Thayasivam(2022)Mohamed and Thayasivam]{9906230}
Isham Mohamed and Uthayasanker Thayasivam.
\newblock Low resource multi-asr speech command recognition.
\newblock In \emph{2022 Moratuwa Engineering Research Conference (MERCon)}, pp.\  1--6, 2022.
\newblock \doi{10.1109/MERCon55799.2022.9906230}.

\bibitem[Murindanyi et~al.(2023)Murindanyi, Yiiki, Katumba, and Nakatumba-Nabende]{murindanyi2023explainable}
Sudi Murindanyi, Brian~Afedra Yiiki, Andrew Katumba, and Joyce Nakatumba-Nabende.
\newblock Explainable machine learning models for swahili news classification.
\newblock In \emph{Proceedings of the 2023 7th International Conference on Natural Language Processing and Information Retrieval}, pp.\  12--18, 2023.

\bibitem[Reshamwala et~al.(2021)Reshamwala, Shah, and Naik]{23Reshamwala2021}
T.~Reshamwala, C.~Shah, and S.~Naik.
\newblock Computerization in home: Change in way of life.
\newblock In \emph{2021 Second International Conference on Smart Technologies in Computing, Electrical and Electronics (ICSTCEE)}, pp.\  1--6, 2021.
\newblock \doi{10.1109/ICSTCEE54422.2021.9708568}.

\bibitem[Tomasello et~al.(2022)Tomasello, Shrivastava, Lazar, Hsu, Le, Sagar, Elkahky, Copet, Hsu, Adi, Algayres, Nguyen, Dupoux, Zettlemoyer, and Mohamed]{tomasello2022stop}
Paden Tomasello, Akshat Shrivastava, Daniel Lazar, Po-Chun Hsu, Duc Le, Adithya Sagar, Ali Elkahky, Jade Copet, Wei-Ning Hsu, Yossi Adi, Robin Algayres, Tu~Ahn Nguyen, Emmanuel Dupoux, Luke Zettlemoyer, and Abdelrahman Mohamed.
\newblock Stop: A dataset for spoken task oriented semantic parsing, 2022.

\bibitem[Yang et~al.(2020)Yang, Yu, and Jia]{commands_CNN}
Xuebin Yang, Hongzhi Yu, and Lei Jia.
\newblock Speech recognition of command words based on convolutional neural network.
\newblock In \emph{2020 International Conference on Computer Information and Big Data Applications (CIBDA)}, pp.\  465--469, 2020.
\newblock \doi{10.1109/CIBDA50819.2020.00110}.

\end{thebibliography}
\bibliographystyle{iclr2024_conference}

\appendix

\end{document}